\documentclass[review]{elsarticle}
\usepackage{lineno,hyperref}
\modulolinenumbers[5]

\journal{Integration: the VLSI Journal}

\usepackage{cite}
\usepackage{amsmath,amssymb,amsfonts}
\usepackage{algorithmic}
\usepackage{graphicx}
\usepackage{textcomp}
\def\BibTeX{{\rm B\kern-.05em{\sc i\kern-.025em b}\kern-.08em
   T\kern-.1667em\lower.7ex\hbox{E}\kern-.125emX}}

\begin{document}
\begin{frontmatter}
\title{MULTI-CHANNEL LOCK-IN BASED DIFFERENTIAL FRONT-END FOR BROADBAND RAMAN SPECTROSCOPY\\}

\author{A. Ragni, G. Sciortino, M. Sampietro, G. Ferrari}
\address{\textit{Dipartimento di Elettronica, Informazione e Bioingegneria} \\
\textit{Politecnico di Milano}, Milano, Italy \\
andrea.ragni@polimi.it}

\author{F. Crisafi, V. Kumar, G. Cerullo, D. Polli}
\address{\textit{Dipartimento di Fisica} \\
\textit{Politecnico di Milano}, Milano, Italy \\
dario.polli@polimi.it}

\begin{abstract}
In Broadband Stimulated Raman Spectroscopy, the intrinsic limit given by the laser shot noise is seldom reached due to the electronic noise of the front-end amplifier and the intensity fluctuations of the laser source. In this paper we present a low-noise multi-channel acquisition system, with an integration-oriented design, able to compensate the common-mode fluctuations of the laser output power with the pseudo-differential structure and reach a sensitivity better than 10 ppm thanks to the lock-in technique.
\end{abstract}


\end{frontmatter}

\section{Introduction}

This work was conducted as part of the ERC-VIBRA project - Very fast Imaging by Broadband coherent RAman - headed by Dario Polli in the Department of Physics of Politecnico di Milano. The goal is to build an innovative microscope for real-time and non-invasive functional characterization of tissues and cells. Final application concerns cancerous cells differentiation and detection of neuronal tumours.

Every organic compound, cell or tissue, is characterized by a specific vibrational spectrum, which is a function of the chemical bonds and structures of molecules. Stimulated Raman spectroscopy (SRS) is commonly used in chemistry to provide that spectrum by which molecules can be identified and analysed. This technique, non-contact and label-free, is based on the Raman scattering effect, an inelastic scattering of photons by molecules discovered in the early 1930s by the Indian scientist C.V. Raman. 

Samples under test interact with two synchronized pulsed lasers, called Pump and Stokes, having different wavelengths in the near-infrared region (NIR). When a Pump photon hits a molecule, this may first raise to a virtual vibrational level and successively relax to a lower level different from the ground state. It results that a photon with lower energy can be emitted at Stokes wavelength (Raman signal or Raman gain). The presence of the latter laser (the Stokes) enhances the relaxation transition because when the Stokes-Pump frequency difference matches a vibrational mode of the molecule, all the molecules in the focal volume are resonantly excited \cite{crisafi2017line}. This translates in a signal enhancement by many orders of magnitude with respect to spontaneous Raman scattering, where only one laser is used \cite{kumar2012balanced}. Consequently, acquisition speed is significantly improved opening new possibilities to the video-rate imaging \cite{vibra}. Since each molecule has a specific Raman spectrum, the sample composition can be analysed by measuring the vibrational behaviour on a wide wavenumber range, typically spanning from 100 $cm^{-1}$ to 3500 $cm^{-1}$. Scientific papers in cancer differentiation show how the spectral regions 800-1800$cm^{-1}$ and 2800-3100$cm^{-1}$ are highly informative, respectively, for gastric \cite{teh2008diagnostic} and brain tumour\cite{zhou2012human}.

In the single-frequency SRS, different vibrational modes of molecules can be investigated by a wavelength time-scan of Stokes or Pump laser. Only narrowband lasers are required, but the whole spectrum measurement is slow because of the laser tuning. In broadband SRS, several vibration modes are excited in the same time by using either the Pump or the Stokes broadband. Supposing the correct detection of each wavelength and a parallel elaboration, this results, in principle, in a faster acquisition of the Raman Spectrum.

In the VIBRA setup a commercial 1.55$\mu$m femtosecond Er:fiber 40MHz-oscillator followed by two erbium-doped fiber amplifiers (EDFAs) is used to generate two coherent pulse train beams: 
\begin{itemize}
	\item the first beam goes through a Periodically Poled Lithium Niobate (PPLN) producing, with the second harmonics generation, the narrowband Pump with $\lambda_p =770nm$ and average power of $\sim$1mW
	\item the second beam passes through a Non Linear fiber (NL fiber) that modifies the wavelength and enlarges the spectrum. A broadband Stokes, consisting in sub-20fs pulses, with $\lambda_s=950\div1050nm$ and average power of $\sim$1mW is produced. 
\end{itemize}

\begin{figure}
	\includegraphics[width=\textwidth, keepaspectratio, clip, angle=0]{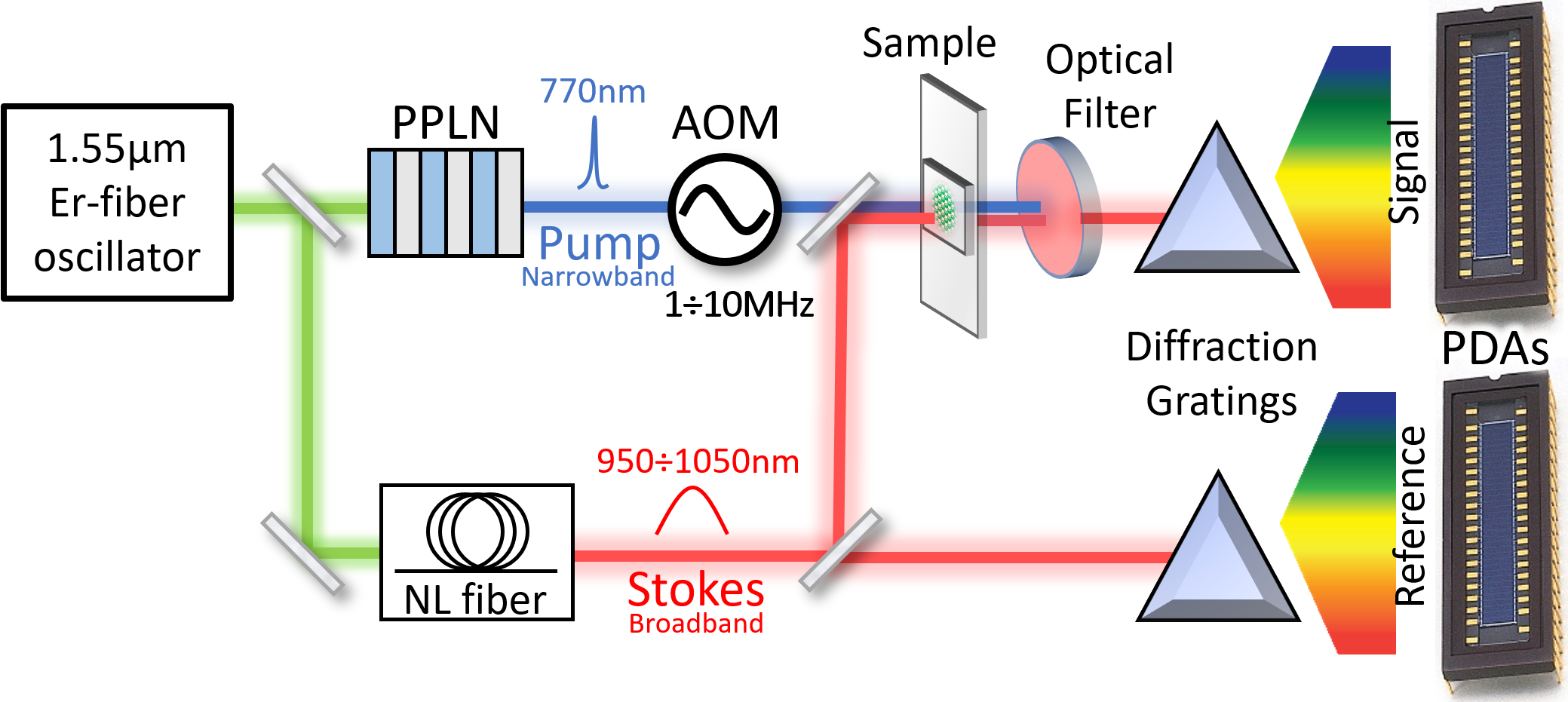}
	\caption{Simplified balanced acquisition scheme adopted in the broadband SRS of the ERC-VIBRA project.}
	\label{fig:vibra_setup}
\end{figure}

The femtosecond laser beam containing the Raman information over a wide range of wavelengths, is spatially diffused on a photodiode array (PDA) with a diffraction grating. Each wavelength of the broadband laser is separately acquired by a specific element of the array and amplified simultaneously with the respective transimpedance amplifier to retrieve the sample Raman spectrum. The Pump beam, having no information, is instead optically filtered. We adopt 32 elements PDA (Osi-optoelectronics A5C-35) because a resolution of 32 intervals in the Raman spectrum is sufficient to differentiate normal cells from cancerous ones. A second PDA is used as reference for a balanced detection in order to compensate the laser common mode noise. Since the detected laser is in the near-infrared range, arrays made of silicon photodiodes provide a sufficient responsivity ($R_{\lambda}\approx$ 0.5A/W) in the whole wavelength range. The VIBRA optical setup is represented in Fig. \ref{fig:vibra_setup}.

The lock-in technique (LIA) has been adopted as detection scheme for two main reasons:

\begin{itemize}
	\item the Stokes beam has an average power $\overline{P_s}$ greater by a factor of $10^4$ or more \cite{freudiger2008label} with respect to the superimposed SRS signal $\Delta P_s$ which has to be measured;
    \item the Stokes noise has a frequency dependence decreasing with $f$ as shown in the following laser noise analisys;
\end{itemize}

This is done by modulating the Pump source in the MHz range ($1\div10$MHz) with an acousto-optic modulator (AOM) before the sample under test. A custom IC with Double-Balanced Mixers is used, at the end of the acquisition chain, to demodulate the Raman signal \cite{ciccarella2016impedance}.

In this paper we address the problem of designing a valid front-end for broadband Raman spectroscopy combining high sensitivity, to detect the weak Raman signal, and high dynamic range to correctly manage the Stokes average power. The latter is limited to few mW (therefore $\sim 100\mu W$ per channel) to avoid the damage of the organic sample during the measurement \cite{talone2018biological}. These requirements together with the multi-channel approach constitute the key elements for an integration-oriented design of the front-end. In order to validate the acquisition scheme before the IC designing process, a first discrete-component prototype with 4 identical differential channels has been designed and tested. A single-frequency SRS experiment on a Methanol sample shows the correct operation of the single channel paving the way to the IC realization.

This paper is an extension of that one originally reported in Proceedings of the 14th International Conference PRIME 2018 \cite{ragni2018lock}. 

\section{Noise sources}
In order to achieve fast Raman frames acquisition, goal of the VIBRA project, the pixel dwell time has to be minimized. This is strictly related to the signal to noise ratio (SNR) that in the lock-in acquisition scheme can be written as:
\begin{equation}
    \frac{S}{N} = \frac{\Delta P_{min}}{RMS_{noise}} =\frac{1}{\sqrt{2}}\cdot \frac{\Delta P_{min}}{\sqrt{\frac{S_n(f)}{4\cdot \tau}}}
    \label{S/N}
\end{equation}
where $\Delta P_{min}$ is the minimum Raman signal, $S_n(f)$ is the total input equivalent noise spectral density and $\tau$ is the low-pass filter time constant. The pixel dwell time is $t_{dwell}\approx 4\cdot\tau$ for a first order low-pass filter.

Typically, in SRS the main noise sources are:
\begin{itemize}
    \item Shot Noise
    \item Laser Intensity Noise 
    \item Electronics Noise
\end{itemize}
The Shot Noise ($S_{shot}$), modelled by a Poisson process, is proportional to the mean power detected by the photodetector ($\overline{P}$) and it represents the theoretical limit for these measurements (eq.\ref{eq:shot}).

\begin{figure}
	\centering\hspace*{-1.3cm}\includegraphics[width=10cm, keepaspectratio, clip, angle=0]{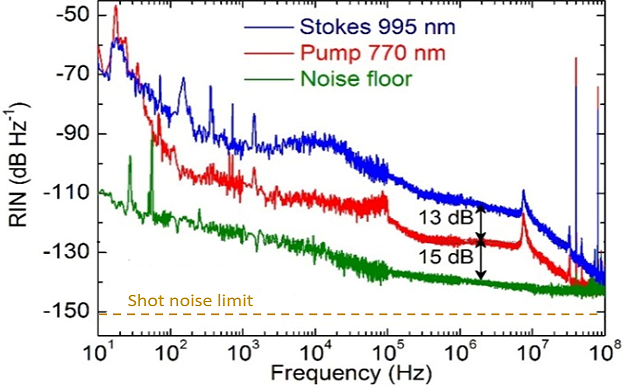}
	\caption{Measured Relative Intensity Noise (RIN) of Stokes (Blue) and Pump (Red) with 1.6mW Pump power \cite{coluccelli2014er}. The Raman signal adds to the Stokes which gives the main contribution to the noise if no compensation technique is adopted (Stokes $RIN\approx-115dB Hz^{-1}$ in the MHz range). \textsuperscript{\textcopyright} 2018 IEEE.}
	\label{fig:RIN}
\end{figure}

The Intensity Laser Noise is due to the power fluctuations and can be the major contribution to the noise in SRS measurements \cite{demtroder2013laser} if not correctly compensated. 
Figure \ref{fig:RIN} shows the Relative Intensity Noise (RIN) of the \textit{Femto Fiber Pro} commercial laser. The RIN represents the noise spectral density of the laser $S_{laser}$ normalized for the average power $\overline{P}$ and is defined as 
\begin{equation}
    RIN = 10\cdot\log{\left(\frac{S_{laser}}{\overline{P^2}}\right)} \left[\frac{dB}{Hz}\right]
    \label{eq:RIN}
\end{equation}
Since the laser noise behaviour decreases with frequency, we decided to modulate the Raman signal in the MHz range, where the noise level is lower.
From Fig.\ref{fig:RIN} is possible to calculate the Stokes intensity noise $S_{Stokes}$ and make a comparison with the theoretical shot noise $S_{shot}$, only depending on the mean power \cite{freudiger2014stimulated}. For an average power of 100$\mu W$ these value are:
\begin{equation}
        S_{Stokes}(f) \approx\left(177.8\frac{pW}{\sqrt{Hz}}\right)^2
        \label{eq:stokes}
\end{equation}
\begin{equation}
        S_{shot}(f)=2\frac{hC_0}{\lambda}\overline{P}\approx\left(6.3 \frac{pW}{\sqrt{Hz}}\right)^2
    \label{eq:shot}
\end{equation}
Where $\lambda\approx1000nm$ and h and $C_0$ are, respectively, the Planck constant and the speed of light in vacuum. Equations \ref{eq:stokes} and \ref{eq:shot} show that intensity fluctuations of the Stokes source are the main contribution to the overall noise. Moreover, they depend on the working conditions (i.e. wavelength, temperature, humidity). A differential architecture has been adopted in this work in order to compensate these Stokes fluctuations being a common mode contribution. The adopted subtractor and mixer are analog circuits to overcome the resolution limit of digital lock-in amplifiers \cite{Carminati2016}.

The last contribution to the noise is given by the Electronics, thus by the acquisition system. The prototype presented in this paper has been designed keeping this term lower than the Shot noise.
%--------------------------------------------------------------------
\section{Analysis of the proposed solution}
\begin{figure*}
	\includegraphics[width=\textwidth, keepaspectratio, clip, angle=0]{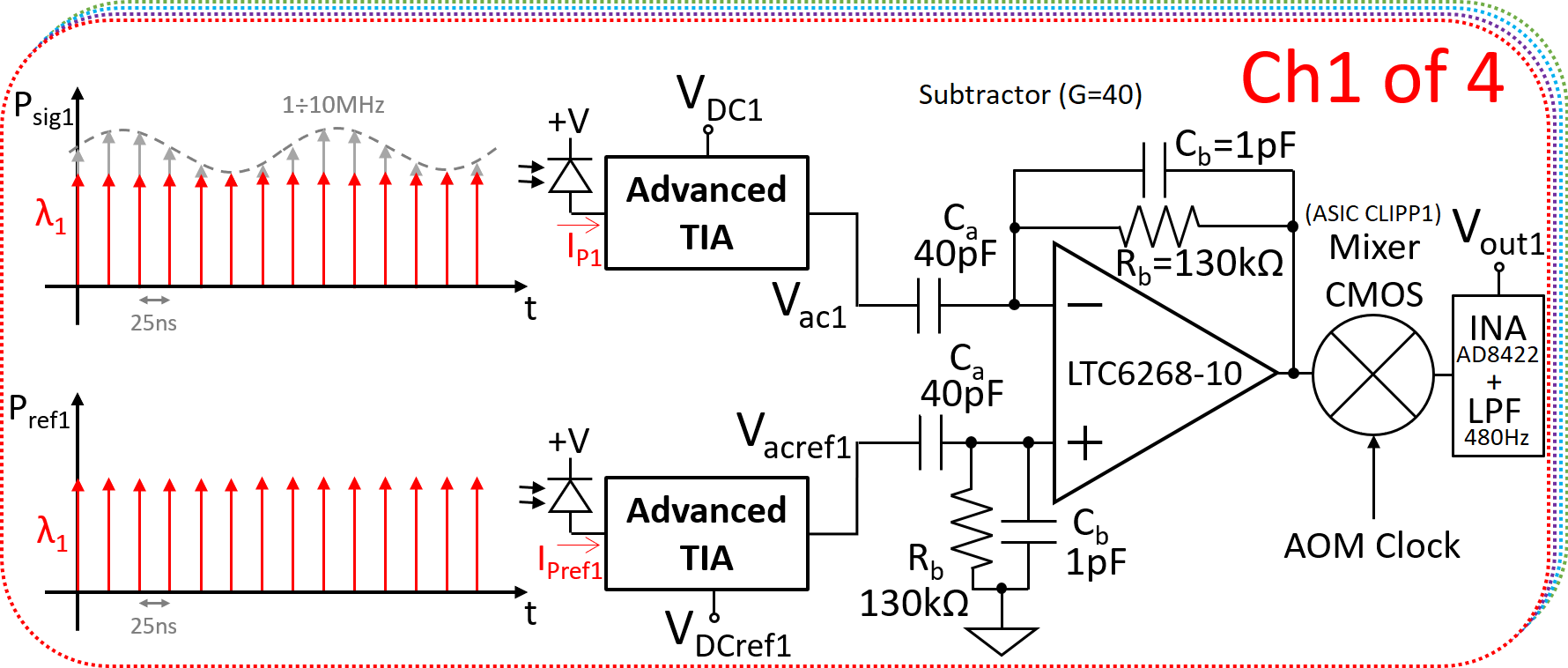}
	\caption{Balanced acquisition scheme of 1 channel over 4 inside the prototype (Ch1 = Red, Ch2 = Purple, Ch3 = Blue and Ch4 = Green). The power graph $P_{sig1}$ is not drawn to scale because the Raman signal (Grey) over the Stokes pulses (Red) is tipically 4 $\div$ 5 orders of magnitude smaller.}
	\label{fig:4-ch}
\end{figure*}

A valid front-end for this application should have the following characteristics:
\begin{itemize}
    \item working with common cathode array of silicon photodiodes with 4$mm^2$ active area and 10pF of junction capacitance (e.g. A5C-35 Osi-Optoelectronics).
    \item handle a low power laser (few tenth of microwatts per channel) as in broadband Raman spectroscopy to avoid sample damaging.
    \item High-bandwidth to amplify the weak Raman signal modulated in the frequency range $1MHz\div10MHz$. The total gain in this interval has to be greater than 10M$\Omega$ to get an appreciable output voltage (mV or V) from the weak Raman signal (10ppm to 1000ppm).
    \item Electronics noise lower than shot noise as seen in the previous section and differential structure for intensity laser noise compensation;
    \item Separated outputs for DC and AC needed for Raman normalization;
    \item Compact structure, lock-in based and multi-channel oriented.
\end{itemize}
Figure \ref{fig:4-ch} shows the Balanced acquisition scheme of the broadband SRS experiment with the 4-ch prototype, lock-in based, developed in this work. Each photodiodes pair of the 2 PDAs detects a single colour of the broadband Stokes which is a 40MHz pulse train. Since this laser has been splitted in two branches, the Raman gain will occur only where the sample is present. The second branch, having the same power fluctuations, is used instead as a reference.

The photodiodes pair of the x-Ch produces the current signals $I_{Px}$ and $I_{Prefx}$ respectively, these are mainly composed by the following terms:
\begin{equation}
I_{Px}= I_{DCx}+i_{trainx}+i_{Snoisex}+\sqrt{i_{shotx}^2}+i_{Ramanx}
\label{eq:Ip}
\end{equation}
\begin{equation}
I_{Prefx}=I_{DCx}+i_{trainx}+i_{Snoisex}+\sqrt{i_{shot,refx}^2}
\label{eq:Ipref}
\end{equation}
  
where $I_{DCx}$ is the current proportional to the average power of the Stokes through the photodiode responsivity ($R_{\lambda}\approx$ 0.5A/W at 1000nm). This term is important to normalize the Raman spectrum, so its value must be acquired before the differential amplifier. The $i_{trainx}$ consists in the AC 40MHz pulsed-train response and $i_{Snoisex}$ represents the common mode fluctuations (equal for both $I_{Px}$ and $I_{Prefx}$). These two terms can be, in principle, cancel out with the balanced detection. Then, $i_{shotx}$ is the shot noise which is uncorrelated between the two beams and finally $i_{Ramanx}$ represents the weak Raman signal ($\sim$ $10^{-5}$ respect to $I_{DC}$) which is only present in the "Signal" branch. The difference between $I_{Px}$ and $I_{Prefx}$ corresponds to the Raman signal, that we want to measure, with the intrinsic noise given by the shot.

\begin{figure}
	\centering\hspace*{0cm}\includegraphics[width=\textwidth, keepaspectratio, clip, angle=0]{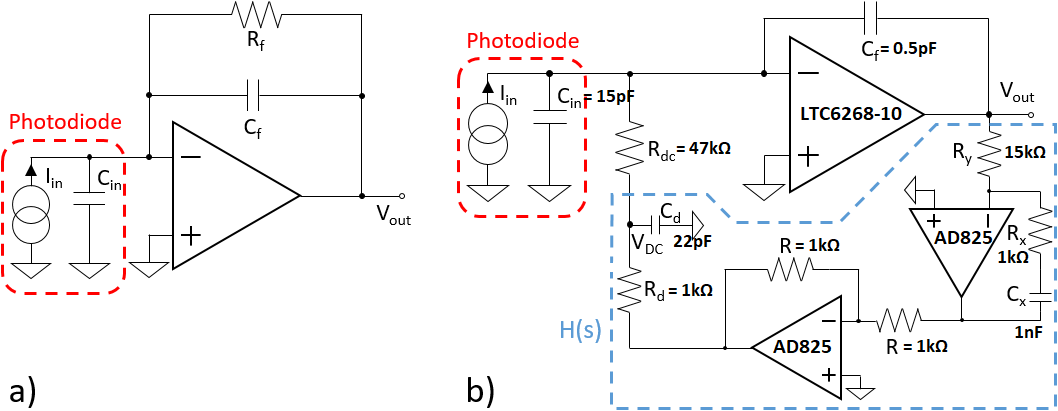}
	\caption{Comparison between Standard TIA configuration (a) and advanced version (b) used as first-stage in this work \cite{ferrari2007wide}. The current source $I_{in}$ models the photodiode while $C_{in}$ represents the sum of all the stray capacitances connected to the virtual ground node of the amplifier, the major contribution is given by the photodiode capacitance $C_{pd} \approx 10pF$. The DC feedback network (with $R_{DC} = 47k\Omega$) in the advanced topology manages the DC current while the Raman signal is amplified with the LTC6268-10 (GBWP=4GHz) having the feedback capacitance $C_f = 0.5pF$. The second stage (subtractor) and the INA have a gain $G_{diff} = 40$ and $G=10$ respectively. The first order low-pass filter at the end of the lock-in scheme has a cut-off frequency $f_c\approx480Hz$ corresponding to $\tau\approx330\mu s$.}
	\label{fig:TIA}
\end{figure}

If a standard TIA is connected to the photodiode (Fig. \ref{fig:TIA}.a), the output voltage in the ideal case is a single-pole transfer function as follow:

\begin{equation}
V_{out}(s)=-\frac{R_f}{1+sR_f C_f}\cdot I_{in}(s)
\label{eq:TF}
\end{equation}

The feedback capacitor $C_f$ should, in principle, be minimized because it sets the TIA bandwidth. However, since the stage $G_{loop}$ is a function of $\frac{C_f}{C_f+C_{in}}$, there is a low limit due to all the parasitic capacitances related to the virtual ground node of the amplifier being included in $C_{in}$. The photodiode capacitance, which is about $10pF$ for an active area of $\approx 4mm^2$ and with a reverse voltage $V_R = $$\sim$$15V$, gives the main contribution. Furthermore, because of stray capacitances are in the order of few hundreds of femtofarad for discrete components technology, working with lower value of $C_f$ is not trivial. In practice, the feedback resistor $R_f$ sets both the DC gain and the bandwidth, so there is a trade-off. Its value should be maximized in order to reduce the input equivalent current noise being:
$$S_n^{eq}(f) \approx \overline{e_n^2} \omega^2 C_{tot}^2 + \frac{\overline{e_n^2}}{R_f^2}+\frac{4kT}{R_f}$$
but this way the bandwidth it is also reduced. Furthermore, since the output voltage $V_{out}$ is proportional to the short and powerful laser train pulses, $R_f$ value is strongly limited by the operational amplifier output dynamic range to prevent clipping.

These limitations are overcome with the advanced TIA topology \cite{ferrari2007wide}, used in this work, having an additional feedback network $H(s)$ needed to manage the DC bias current (see Fig. \ref{fig:TIA}.b). The network $H(s)$ in series to the resistance $R_{DC}$ has the expression:
$$H(s) = \frac{V_{dc}(s)}{V_{out}(s)} = -\frac{1}{R_y}\cdot\frac{1+sR_xC_x}{sC_x}\cdot\frac{1}{1+sR_dC_d}$$

Therefore, the transfer function H(s) is designed to have high gain in DC and high attenuation in the signal bandwidth. The AC part of $I_{Px}$, including the pulse-train and the Raman signal, results amplified by the integrator at the output node:
\begin{equation}
V_{out}(s)=-\frac{1}{sC_f}\cdot i_{ac}(s)
\label{eq:Vout1}
\end{equation}

Having and integrator transfer function, the 40MHz unwanted component of the laser is attenuated, with respect to the signal, and this improves the dynamic-range.
Differently, the DC component $I_{DC}$ of the photodiode (given by the Stokes average power) is made to flow through the resistor $R_{DC}$ by the $H(s)$ network. This results in the voltage:
\begin{equation}
V_{DC}(s)=-R_{DC}\cdot I_{DC}(s)
\label{eq:Vdc}
\end{equation}

The advanced TIA topology allows to strongly relax the amplifier output dynamics because the two components of the current $I_{Px}$, AC and DC, are amplified on different nodes. Moreover, the operational amplifiers in the $H(s)$ can have a wide power supply voltage because they operate at low frequency. This translates in a greater $R_{DC}$ that will benefit the input noise as shown in the equation \ref{equ:noise}. The operational amplifier in the integrator stage can only be optimized for bandwidth and noise performance because there is no more the gain-bandwidth trade-off seen for the standard TIA. The bandwidth is proportional to the O.A. gain-bandwidth product through the factor $\frac{C_f}{C_f+C_{in}}$ and it does not depend on the $R_{DC}$ value. With the operational amplifier LTC6268-10 (GBWP=4GHz), selected for this work, the bandwidth is $\sim$100MHz. Finally, with this advanced TIA topology, the information about the Stokes average power $\overline{P_s}=-V_{DC}/R_{\lambda}$, useful for the Raman signal normalization $\frac{\Delta P_s}{\overline{P_s}}$, is already available on the $V_{DC}$ node.

The input equivalent current noise of the advanced TIA is:
\begin{equation}
\label{equ:noise}
\bar{i}^2_{neq} \approx \bar{i}^2_{n} +  \bar{e}^2_{n}\omega^2\left(C_f+C_{in}\right)^2 + \frac{\bar{e}^2_{n}}{R^2_{DC}} +\frac{\bar{e}^2_{H(s)}}{R^2_{DC}}+\frac{4kT}{R_{DC}} 
\end{equation}
where $\bar{i}^2_{n}\approx(7 \frac{fA}{\sqrt{Hz}})^2$ and $\bar{e}^2_{n}\approx(4 \frac{nV}{\sqrt{Hz}})^2$ are the O.A. current and voltage equivalent input noise, respectively, and $\bar{e}^2_{H(s)}$ is the voltage equivalent noise of the $H(s)$ network. By using a FET operational amplifier and $R_{DC} > 1k\Omega$, at the signal frequency of 1 MHz, the first stage input referred noise can be approximated with the following expression:
\begin{equation}
\label{equ:noise2}
\bar{i}^2_{neq} \approx \bar{e}^2_{n}\omega^2\left(C_f+C_{in}\right)^2 +\frac{4kT}{R_{DC}} 
\end{equation}

In the modulation range, this equation shows that the input referred current noise increases with frequency but thanks to the integrator transfer function of the input stage, it becomes white before entering the CMOS mixer. Therefore, by demodulating with a square-wave and considering all the harmonics, the output noise is not amplified.
In the frequency range $1MHz\div10MHz$ and considering an input capacitance of $\approx$ 15 pF and $R_{DC}= 20 k\Omega$, the input equivalent current noise $i^2_{neq}$ goes from $(850 \frac{fA}{\sqrt{Hz}})^2$ to $(5 \frac{pA}{\sqrt{Hz}})^2$. In principle, these values are of the same order of the Stokes shot noise (eq.\ref{eq:shot}) divided by the squared responsivity, $R_{\lambda}\approx$ 0.5A/W in the NIR, making possible a shot noise limited measurement using this front-end.  
The capacitance and resistors values have been selected to be easily integrated in an ASIC using a standard 3.3V CMOS 0.35$\mu m$ process, with poly-poly capacitance module and high resistance polysilicon module. Furthermore, the signal is modulated in the MHz range, far above the $\frac{1}{f}$ noise corner frequency of the nMos and pMos transistors.

\begin{figure}
	\centering\hspace*{0cm}\includegraphics[width=\textwidth, keepaspectratio, clip, angle=0]{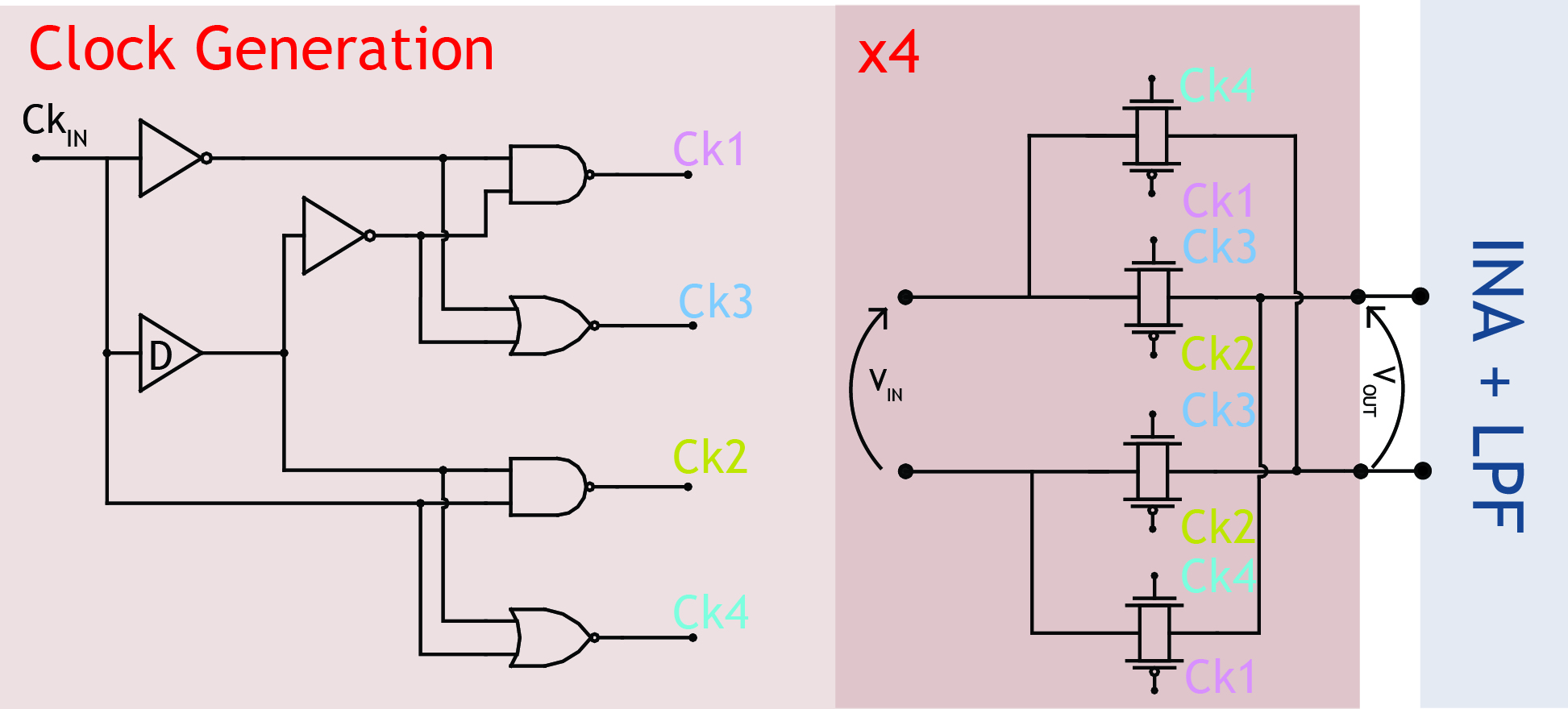}
	\caption{Internal scheme of the custom multichannel mixer. Non-overlapped clock generation circuit (on the left). Double-balanced switching scheme (on the right).}
	\label{fig:mixer}
\end{figure}

For each channel, a couple of advanced TIAs are connected to a second stage differential amplifier with gain $G_{diff} = 40$ as shown in Fig. \ref{fig:4-ch}. This stage can be easily implemented in a custom ASIC using a fully differential stage.
The common mode laser fluctuations are compensated by the balanced acquisition depending on the CMRR (Common-mode rejection ratio) of the stage and, in principle, only the Raman signal and the uncorrelated noise sources (as shot and thermal noise) get amplified. 
With a custom ASIC having 4 double-balanced mixers, the signal coming from each channel is then demodulated to baseband and finally amplified and filtered with an instrumentation amplifier (INA with $G=10$) followed by a low-pass filter (LPF with $f_c\approx480Hz$). The chip is a simple structure made by 4 transmission gates switches per each channel as shown in Fig. \ref{fig:mixer}, that are driven by the clock signal used by the AOM to modulate the Pump.
In order to have fast rising edges, the external clock signal goes through a digital on-chip stage, that generates four clock signals non-overlapped in phase and in counter-phase. An external all-pass filter, finally, is used to control the clock phase in order to maximize the output signal.

\section{Experimental validation and discussion}
\begin{figure}
	\includegraphics[width=\textwidth, keepaspectratio, clip, angle=0]{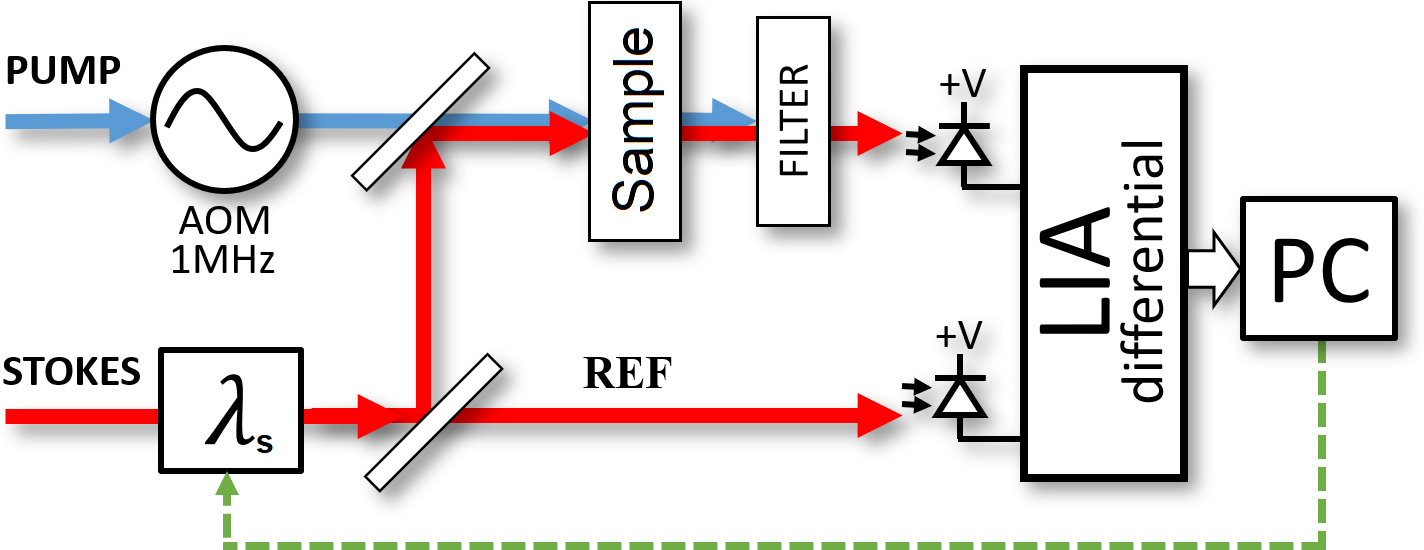}
	\caption{Single-frequency stimulated Raman spectroscopy with balanced acquisition. The whole Raman spectrum is acquired with a time-scan using a single channel of the prototype. The Stokes wavelength is controlled with a PC over the interval $950nm\div1050nm$.}
	\label{fig:scheme}
\end{figure}

To characterize and experimentally validate the acquisition scheme developed in the proposed front-end, a single-frequency Raman spectroscopy experiment has been carried out on a liquid phase Methanol sample (MeOH) in the C-H stretch region using one channel of the prototype. A simplified scheme of the adopted setup is represented in Fig.\ref{fig:scheme}. The Stokes power was equal to $P_s\approx40\mu W$ for one channel to simulate the 32 channels broadband SRS experiment where the total power is limited to avoid the organic sample damaging. That is the average power acquired by the photodiode and measured from the DC output voltage of the front-end, thus keeping in account the responsivity $R_{\lambda}\approx$ 0.5A/W and the DC gain $R_{dc}=47k\Omega$. The modulation frequency was chosen to $f_m=1MHz$ to make the front-end working in the optimal operating point for minimum noise. In fact, in this condition the input equivalent power noise is equal to $S_{el}\approx (2.4 \frac{pW}{\sqrt{Hz}})^2$ (considering the photodiode responsivity $R_{\lambda}\approx$ 0.5A/W in the NIR) which is lower than Stokes shot noise $S_{shot}\approx (7.2 \frac{pW}{\sqrt{Hz}})^2$ with $P_s\approx40\mu W$. Note that the input equivalent power noise includes a factor 2 given by the sum, in the signal and reference path, of the uncorrelated terms: Shot and Electronics noise.

Referring to Fig. \ref{fig:4-ch}, the output voltages of the first channel: 
\begin{itemize}
    \item $V_{out1}$ containing the Raman information
    \item $V_{DC1}$ and $V_{DCref1}$ for the power balancing and normalization
\end{itemize}
were sampled with a data acquisition system (NI USB-6259 by National Instruments) and sent to a PC. A NI LabVIEW interface was used to control the laser power and perform the Stokes wavelength time-scan in the interval 950nm to 1050nm. Stokes power in the two branches is manually adjusted with a polarizer and a neutral density filter.

\begin{figure}
	\centering\hspace*{-0.4cm}\includegraphics[width=10.4 cm, keepaspectratio, clip, angle=0]{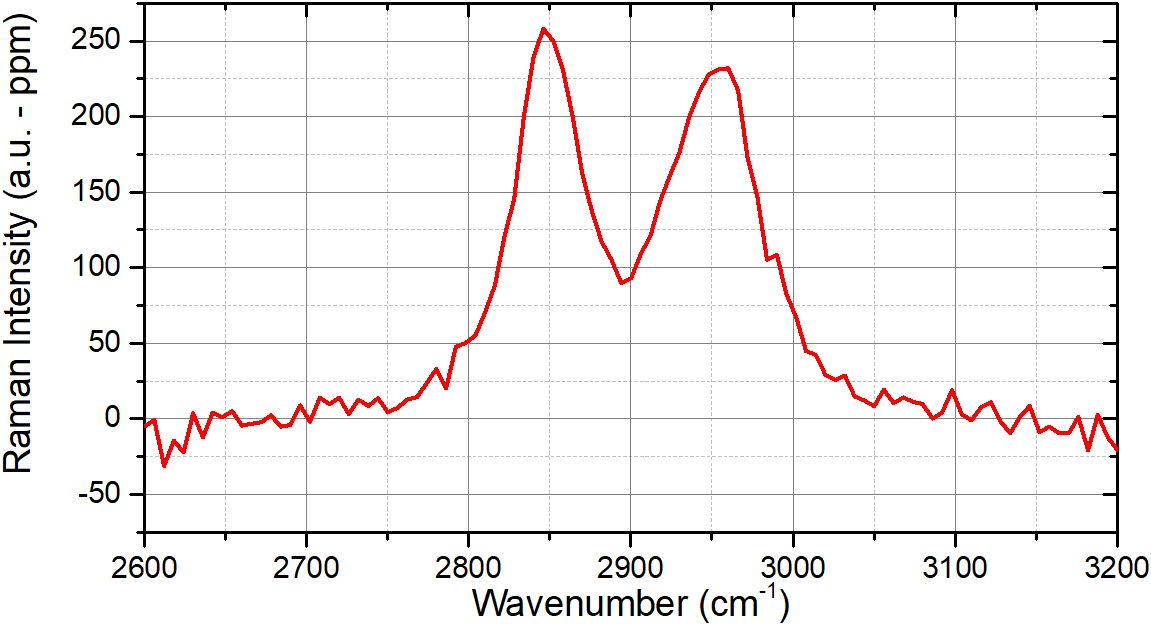}
	\caption{Normalized Raman spectrum $(\Delta P_s/P_s)$ of Methanol (MeOH) obtained with the proposed amplifier operating with $\tau = 330\mu s$ and with $\sim$$40\mu W$ Stokes power. The obtained Methanol spectrum can be compared with those ones of papers \cite{rehault2015broadband} and \cite{yu2013complete}. \textsuperscript{\textcopyright} 2018 IEEE.}
	\label{fig:methanol}
\end{figure}

The normalized Raman spectrum of the Methanol sample, plotted in Fig. \ref{fig:methanol}, was correctly acquired by the proposed front-end with a filter time constant $\tau=330\mu s$. The spectrum exhibits two peaks typical of C-H stretching region, respectively $\sim$2850$cm^{-1}$ for symmetric stretching and $\sim$2950$cm^{-1}$ for anti-symmetric stretching \cite{rehault2015broadband}, \cite{yu2013complete}.

\begin{figure}
	\centering\hspace*{0cm}\includegraphics[width=10 cm, keepaspectratio, clip, angle=0]{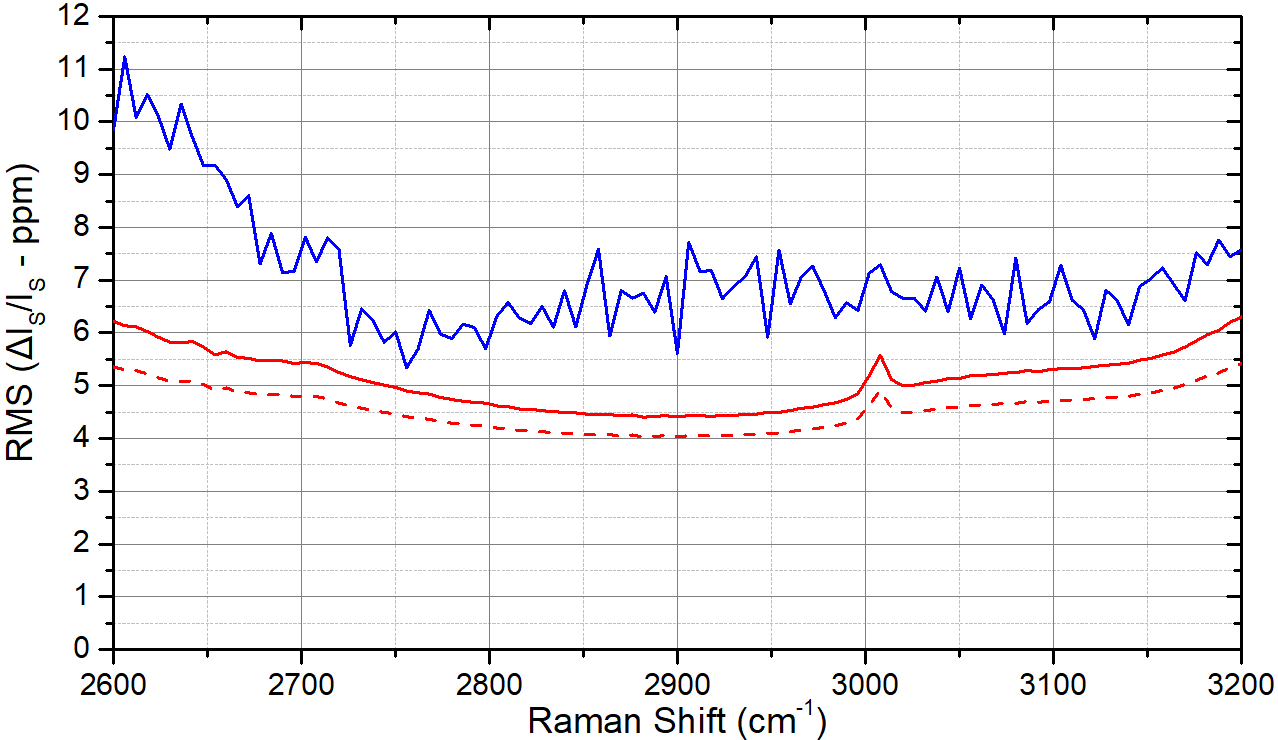}
	\caption{Normalized RMS noise spectra: measured (blue), theoretical noise (red) which includes shot noise (dotted red) and electronics noise. The blue line also indicates the minimum normalized Raman intensity detectable with SNR=1. \textsuperscript{\textcopyright} 2018 IEEE. }
	\label{fig:rms}
\end{figure}

The experimental results also indicates that the measured noise is almost equal to the theoretical Stokes shot noise, as shown in Fig. \ref{fig:rms}, leading to a sensitivity lower than 10ppm (with SNR=1). This demonstrates the ability of the proposed differential structure to strongly compensate Stokes intensity fluctuations. As further confirmation of this characteristic, a subsequent measurement has indicated a $CMRR\approx56$ at 1MHz, mainly limited by electronics mismatches, enough to make noise of Eq. \ref{eq:stokes} negligible with respect to the intrinsic limit given by the shot noise. This result can be compared with those ones of papers \cite{liao2015microsecond} and \cite{slipchenko2012heterodyne} where a resonance circuit has been used instead of the lock-in technique.

The maximum SNR (Signal-to-Noise Ratio) that we achieved on the two Raman peaks (250ppm Raman intensity), even with an intentionally low laser power ($P_s\approx40\mu W$), can be estimated as follow:
\begin{equation}
    \frac{S}{N} =\frac{1}{\sqrt{2}}\cdot \frac{250ppm\cdot P_{s}}{\sqrt{\frac{S_{shot}}{4\cdot \tau}}} \approx 350
    \label{S/N}
\end{equation}

\section{Conclusions}
In this paper, a novel multi-channel differential lock-in amplifier, for broadband Raman Spectroscopy applications, is presented and tested. The spectrum of liquid phase Methanol is correctly acquired with a single-frequency SRS, using a time constant $\tau\approx330\mu s$, and normalized over the average Stokes power. The resulting SNR is equal to 350 on the Raman peak of the spectrum. The measurement is practically shot noise limited, even with an intentionally low power Stokes $P_s\approx 40\mu W$, thanks to the low-noise advanced TIA - $S_{el}\approx (3.4 \frac{pW}{\sqrt{Hz}})^2$ at $f_m=1MHz$ - and to the balanced acquisition with $CMRR\approx56$. The Stokes power acquired by each photodiode is kept low to simulate the 32 channels broadband SRS experiment where the total power is limited (few mW) to avoid the organic sample damaging. The differential front-end, replicated on each channel of the prototype, has been experimentally validated obtaining a sensitivity lower than 10ppm. The proposed topology can be used with common cathode photodiodes arrays making ideal for broadband Raman Spectroscopy where a multi-channel architecture is needed. Moreover, as long as the Stokes laser is correctly detected by the photodiode with a sufficient responsivity, this front-end structure can be potentially used in every spectral region (100-3500$cm^{-1}$). The advanced TIA and the differential second stage has been designed with discrete components, but with a view to integration. Indeed, a custom integrated circuit with 4 channels based on the same structure has been already designed with AMS CMOS 0.35$\mu$m technology. To better compensate any optical and electronics mismatch, an innovative auto-balance feedback network has been introduced inside the chip for each channel. The experimental results obtained with this novel ASIC will be presented soon.

\section*{Acknowledgment}
We acknowledge support by the European Research Council Consolidator Grant VIBRA (ERC-2014-CoG No. 648615) - Very fast Imaging by Broadband coherent RAman - headed by Dario Polli in the Department of Physics of Politecnico di Milano. A special thanks also goes to Francesco Crisafi and Vikas Kumar for the design and preparation of the optical setup used in the single-frequency experiment.

\bibliography{Bibliografia} 
\end{document}